\documentclass[12pt]{article}
\usepackage{epsf}
\usepackage{amsmath}
\usepackage{graphics}
\usepackage{cite}

\renewcommand{\thefootnote}{\#\arabic{footnote}}
\begin{document}

\newcommand{\gtrsim}{ \mathop{}_{\textstyle \sim}^{\textstyle >} }
\newcommand{\lesssim}{ \mathop{}_{\textstyle \sim}^{\textstyle <} }

\newcommand{\rem}[1]{{\bf #1}}

\renewcommand{\thefootnote}{\fnsymbol{footnote}}
\setcounter{footnote}{0}
\begin{titlepage}

\def\thefootnote{\fnsymbol{footnote}}

\begin{center}
\hfill hep-ph/0610116\\
\hfill October 2006\\
\vskip .5in
\bigskip
\bigskip
{\Large \bf Proton Decay in Teravolt Unification}

\vskip .45in

{\bf P.H. Frampton}

{\em University of North Carolina, Chapel Hill, NC 27599-3255, USA}

\end{center}

\vskip .4in
\begin{abstract}

Model building based on abelian quiver gauge theories gives models 
which resemble trinification, originally proposed with yottavolt unification. 
However, unification occurs in the teravolt range so proton decay
must be completely excluded in the scalar sector. It is straightforward
to accomplish this by a discrete symmetry which is a generalized baryon number B.
Unlike in trinification, this is possible 
because quarks and leptons
acquire masses from relevant operators which appear in four-dimensional
conformal symmetry breaking. 
\end{abstract}
\end{titlepage}

\renewcommand{\thepage}{\arabic{page}}
\setcounter{page}{1}
\renewcommand{\thefootnote}{\#\arabic{footnote}}

\newpage

\bigskip

One approach to the hierarchy, or naturalness, problem is to postulate
conformality, four-dimensional conformal invariance at high energy, for
the non gravitational extension of the standard model.
The conformality
approach suggested\cite{PHF1998} in 1998 has made considerable progress. 
Models which contain the standard model fields have been constructed\cite{Z7}
and a model which grand unifies at about 4 TeV\cite{Z12} has been
examined. The latter is the subject of this note.

The original speculation \cite{PHF1998}
that such models may be conformal has been refined to
exclude models which contain scalar fields
transforming as adjoint representations because only if all
scalars are in bifundamentals are there chiral fermions and,
also only if all scalars are in bifundamentals, the
one-loop quadratic divergences cancel in the scalar propagator.
We regard it as encouraging that these two desirable
properties select the same subset of models.

Another phenomenological encouragement stems from the
observation\cite{Vafa} that the standard model representations for
the chiral fermions can all be accommodated in bifundamentals
of $SU(3)^3$ and can appear naturally in the conformality
approach. The model building yields theories similar to
the trinification proposed in \cite{DGG}. There are three
significant differences between trinification and conformailty:

\begin{itemize}

\item Unification in conformality occurs in the
TeV (Teravolt) range while in trinification it was
in the $10^{15} GeV$ (Yottavolt) range.

\item In trinification the gauge group was quasi-simple
 $SU(3) \times SU(3) \times SU(3) \times Z_3$ with one
unique gauge coupling while in conformality the $Z_3$
is absent.

\item Most importantly, the fermions in conformality
acquire mass from relevant operators which appear in
four-dimensional confromal symmetry breaking, not only
from Yukawa coupling to the Higgs scalar.

\end{itemize}

In the present note we address the issue of proton decay.
Under $SU(3)_C \times SU(3) \times SU(3)$, the quarks and leptons
families each appear in the representations
\begin{equation}
[(3, 3^*, 1) + (1, 3, 3^*) + (3^*, 1, 3)]
\end{equation}
for which the baryon numbers are respectively $B=+1/3, 0, -1/3$
for quarks, leptons and antiquarks.

The gauge bosons cannot transform quarks
into leptons or vice versa because of the factoring out of the
color $SU(3)_C$ group. So unlike in $SU(5)$(\cite{GG}) proton
decay is absent in the gauge sector.

The scalars are likewise in 27's according to
\begin{equation}
[(3, 3^*, 1) + (1, 3, 3^*) + (3^*, 1, 3)] + c.c.
\end{equation}
although here, unlike for the fermions the complex conjugate
representations must be included.

\noindent Fermion masses arise in trinification \cite{DGG} from Yukawa couplings of the form
\begin{equation}
(3, 3^*, 1)_q (3^*, 1, 3)_q (1, 3, 3^*)_{\phi}
\end{equation}
for the quarks and
\begin{equation}
(1, 3, 3^*)_f (1, 3, 3^*)_f (1, 3, 3^*)_{\phi}
\end{equation}
for the leptons. Because these are two independent couplings,
trinification has the feature of giving no relationship between
quark and lepton masses.

There are additional Yukawa couplings possible which would violate
baryon number $B$ and cause catastrophically rapid proton decay
with a TeV unification scale. Therefore these must be forbidden,
as is achieved by assigning a generalization of baryon
number $B=+1/3, 0 -1/3$ respectively to the three representation
listed. 
\begin{equation}
(3, 3^*, 1)_{\phi} ~~~~~~~~~~~~~~~~ B=1/3  
\end{equation} 
\begin{equation}
(1, 3, 3^*)_{\phi} ~~~~~~~~~~~~~~~  B=0 
\end{equation}
\begin{equation}
(3^*, 1, 3)_{\phi} ~~~~~~~~~~~~~~~ B=-1/3 
\end{equation}
Such assignmnents are very natural.

In phenomenological analysis of trinification\cite{Pakvasa},
however, such a procedure is avoided in order to be able
to obtain acceptable
quark and lepton mass matrices. With teravolt unification
such departure from the simplest case
is not possible as proton decay must be totally forbidden.

However, the quark and lepton masses in conformality will
acquire contributions in relevant operators $m_i\psi\psi$
from breaking of conformal symmetry. The pattern of these
masses cannot yet be calculated, although a constraint on
the pattern of the corresponding scalar masses has been
suggested in \cite{DF} based on the cancellation of
quadratic divergences.

In grand unification based on conformality, therefore,
it is expected that proton decay will be completely
absent in the non-gravitational theory. Gravitational
effects may eventually destabilize the proton but with 
lifetime $\sim 10^{50} y$ far beyond any forseeable experiment.

Equally or more important is the realization that consistency
of the conformality with proton decay dictates that the
quark and lepton masses receive significant contributions
from the four-dimensional conformal symmetry breaking, not only
from the Yukawa coupling to Higgs as universally assumed in previous
studies of grand unification. 

In the standard model, 
while the $Z^0, W^{\pm}$ masses are accurately
predicted by a Higgs mechanism, there is no
similar statement about fermion masses. 
My conclusion is that quark and lepton masses arise
principally as relevant operators
\footnote{One important question is how such mass
terms are induced above the electroweak scale. One possibility is via four fermion 
operators. I thank Edoardo
Di Napoli for discussion.} arising from breaking
of four-dimensional conformal invariance; a corollary is 
that the couplings of the standard Higgs scalar to
quarks and leptons depart from the values usually assumed.

\vspace{2.0cm}

This work was supported in part by the
U.S. Department of Energy under Grant No. DE-FG02-06ER-41418.

\end{document}